\documentclass[amsmath,amssymb,aps,twocolumn,pra]{revtex4-1}

\usepackage{ifthen}
\newboolean{NP}
\setboolean{NP}{false}
\newboolean{FIG}
\setboolean{FIG}{false}

\usepackage{amsmath,amssymb}
\usepackage{graphicx}
\usepackage{times}
\usepackage{soul}
\usepackage[normalem]{ulem}
\usepackage[usenames]{color}

\renewcommand{\Re}{{\rm Re}}
\renewcommand{\Im}{{\rm Im}}
\newcommand{\id}{\mathbb{I}}  
\newcommand{\EE}{\mathbb{E}}

\renewcommand{\vec}[1]{\mathbf{#1}}

\ifthenelse{\boolean{NP}}{
\newcommand{\figtitle}[1]{\textbf{#1}}
\bibliographystyle{naturemag}
\newcommand{\ws}{}
}{
\newcommand{\figtitle}[1]{#1}
\newcommand{\ws}{ }
}

\begin{document}
\title{Probing the relaxation towards equilibrium in an isolated strongly correlated 1D Bose gas}

% --- AUTHORS ---
\author{S.\ Trotzky$^{1-3}$, Y.-A.\ Chen$^{1-3}$ , A.\ Flesch$^{4}$, I.\ P.\ McCulloch$^{5}$, U.\ Schollw{\"o}ck$^{1,6}$, J.\ Eisert$^{6,7}$ and I.\ Bloch$^{1-3}$}

% --- AFFILIATIONS ---
\newcommand{\affqmulmu}{Fakult\"at f\"ur Physik, Ludwig-Maximilians-Universit\"at, 80798 M\"unchen, Germany}
\newcommand{\affqmumpq}{Max-Planck Institut f\"ur Quantenoptik, 85748 Garching, Germany }
\newcommand{\affqmumainz}{Institut f\"ur Physik, Johannes Gutenberg-Universit\"at, 54099 Mainz, Germany}
\newcommand{\affjuelich}{Institute for Advanced Simulation, Forschungszentrum J{\"u}lich, 52425 J{\"u}lich, Germany}
\newcommand{\affbrisbane}{School of Physical Sciences, The University of Queensland, Brisbane, QLD 4072, Australia}
\newcommand{\affiasberlin}{Institute for Advanced Study Berlin, 14193 Berlin, Germany}
\newcommand{\affpotsdam}{Institute of Physics and Astronomy, University of Potsdam, 14476 Potsdam, Germany}

\ifthenelse {\boolean{NP}}
% NATURE PHYSICS STYLE
{\maketitle
\begin{affiliations}
\item \affqmulmu
\item \affqmumpq
\item \affqmumainz
\item \affjuelich
\item \affbrisbane
\item \affiasberlin
\item \affpotsdam
\end{affiliations}
}
% REVTEX-STYLE
{ \affiliation{$^1$ \affqmulmu}
\affiliation{$^2$ \affqmumpq}
\affiliation{$^3$ \affqmumainz}
\affiliation{$^4$ \affjuelich}
\affiliation{$^5$ \affbrisbane}
\affiliation{$^6$ \affiasberlin}
\affiliation{$^7$ \affpotsdam}
}

%\date{\today}

\begin{abstract}
The problem of how complex quantum systems eventually come to rest lies at the heart of statistical mechanics. The maximum entropy principle put forward in 1957 by E.\ T.\ Jaynes suggests what quantum states one should expect in equilibrium but does not hint as to how closed quantum many-body systems dynamically equilibrate. A number of theoretical and numerical studies accumulate evidence that under specific conditions quantum many-body models can relax to a situation that locally or with respect to certain observables appears as if the entire system had relaxed to a maximum entropy state. In this work, we report the experimental observation of the non-equilibrium dynamics of a density wave of ultracold bosonic atoms in an optical lattice in the regime of strong correlations. Using an optical superlattice, we are able to prepare the system in a well-known initial state with high fidelity. We then follow the dynamical evolution of the system in terms of quasi-local densities, currents, and coherences. Numerical studies based on the time-dependent density-matrix renormalization group method are in an excellent quantitative agreement with the experimental data. For very long times, all three local observables show a fast relaxation to equilibrium values compatible with those expected for a global maximum entropy state. We find this relaxation of the quasi-local densities and currents to initially follow a power-law with an exponent being significantly larger than for free or hardcore bosons. For intermediate times the system fulfills the promise of being a dynamical quantum simulator, in that the controlled dynamics runs for longer times than present classical algorithms based on matrix product states can efficiently keep track of.
\end{abstract}

\ifthenelse {\boolean{NP}}
{}{\maketitle}

% --- FIGURE CAPTIONS ---
\newcommand{\captionOne}{\figtitle{Relaxation of the density pattern.} (a) Concept of the experiment: after having prepared the density wave $|\psi(t=0)\rangle$ ({\it i}), the lattice depth was rapidly reduced to enable tunneling ({\it ii}). Finally, the properties of the evolved state were read out after all tunneling was suppressed again ({\it iii}). (b) Even-odd resolved detection: particles on sites with odd index were brought to a higher Bloch band. A subsequent band-mapping sequence was used to reveal the odd- and even-site populations\ws\cite{Sebby:2006,Foelling:2007a}. (c) Integrated band-mapping profiles versus relaxation time $t$ for $h/(4J) \simeq 0.9\,{\rm ms}$, $U/J = 5.16(7)$ and $K/J \simeq 9 \times 10^{-3}$. (d) Odd-site density extracted from the raw data shown in c. The shaded area marks the envelope for free Bosons (light grey) and including inhomogeneities of the Hubbard parameters in the experimental system (dark grey).}

\newcommand{\captionTwo}{\figtitle{Relaxation of the local density for different interaction strengths.} We plot the measured traces of the odd-site population $n_{\rm odd}(t)$ for four different interaction strengths $U/J$ (circles). The solid lines are ensemble-averaged results from $t$-DMRG simulations without free parameters.
The dashed lines represent simulations including next-nearest neighbor hopping with a coupling matrix element $J_{\rm NNN}/J \simeq 0.12$ (a), $0.08$ (b), 0.05 (c) and 0.03 (d) calculated from the single-particle band structure. 
}

\newcommand{\captionThree}{\figtitle{Quasi-local current measurement.} (a) To measure the quasi-local density flow every second tunnel coupling was suppressed, coupling either odd-even or even-odd pairs. (b) Oscillations of the odd-site population in the double-wells with fitted sine-waves for $t = 100\,{\rm \mu s}$ (solid), $200\,{\rm \mu s}$ (dashed) and $400\,{\rm \mu s}$ (dotted). The value of $U/J$ during the relaxation was $5.16(7)$. (c) Extracted amplitude $A$ and phase $\phi$ of the double-well oscillations for odd-even (filled circles) and even-odd (open circles) couplings. The solid lines show the respective results of the $t$-DMRG simulations. The dashed lines are fits to a linear increase in the phase and a power-law decay of the amplitude. The insets show the amplitude in a log-log plot (left) and the extracted power-law coefficients (right). The horizontal grey line indicates the power-law coefficient $\alpha = 0.5$ for free and hardcore bosons.}

\newcommand{\captionFour}{\figtitle{Build-up of short-ranged correlations.} (a) Plot of the integrated density profiles obtained after ToF versus $4Jt/h$ for $U/J = 5.16(7)$ as obtained in the experiment (left) and reconstructed from numerical $t$-DMRG simulations (right). The images show the crossover from a purely Gaussian distribution ($t = 0$) to a more complex quasi-momentum distribution ($0< 4Jt/h < 2$) to a purely sinusoidal pattern ($4Jt/h>2$). (b) Visibility of the interference patterns versus $4Jt/h$ obtained experimentally (circles) and from the simulations (solid curve). The grey line represents the measured visibility at $4Jt/h \simeq 5$, while the dashed line corresponds to the value obtained from the simulation of a homogeneous system\ws\cite{Flesch:2008}. (c) Steady-state value of the visibility measured at $4Jt/h \simeq 5$. The solid line is a guide for the eye $\propto J/U$.}

% --- ACKNOWLEDGEMENTS ---
\newcommand{\thankyou}{We acknowledge stimulating discussions with B.~Paredes, M.~Cramer and C.~Gogolin. This work was supported by the DFG (FOR 635, FOR 801), the EU (NAMEQUAM, QESSENCE, MINOS, COMPAS), the EURYI, and DARPA-OLE.}

% --- AUTHOR CONTRIBUTIONS ---
\newcommand{\contributions}{J.~E., U.~S. and I.~B. conceived the research. S.~T. and Y.-A.~C. performed the experiments and evaluated the data. A.~F., U.~S. and I.~McC. set up the $t$-DMRG code and carried out the time-dependent numerical simulations. J. E. performed the analytical calculations and the analysis of mean-field and Markovian approaches. All authors discussed the results and wrote the manuscript.}

% --- ADDITIONAL INFORMATION ---
\newcommand{\declare}{The authors declare that they have no competing financial interests. Correspondence and requests for materials should be addressed to S.~T.~(email: stefan.trotzky@lmu.de) or A.~F.~(email: a.flesch@fz-juelich.de).}

Ultracold atoms in optical lattices provide highly controllable quantum systems allowing to experimentally probe various quantum many-body phenomena. In this way, ground state properties of Hamiltonians that play a fundamental role in the condensed matter context have been investigated under precisely tunable conditions\ws\cite{Jaksch:2005, Lewenstein:2007, Bloch:2008}. Features that are even harder to probe in actual condensed matter materials or to simulate in numerical studies are dynamical ones, including dynamical properties emerging in adiabatic sweeps\ws\cite{Chen:2010} and far from equilibrium\ws\cite{Polkovnikov:2010,Greiner:2002a, Sebby:2007a, Will:2010a, Tuchman:2006, Fertig:2005, StamperKurn:2006}. In this respect, for example, the quench from a shallow to a deep optical lattice\ws\cite{Greiner:2002a,Sebby:2007a,Will:2010a} and the phase dynamics emerging after splitting a one-dimensional Bose liquid\ws\cite{Hofferberth:2007} have previously been studied experimentally.

In this article, we report on the direct observation of relaxation dynamics in an interacting
many-body system using ultracold atoms in an optical lattice. Starting with a patterned density with alternating empty and occupied sites in isolated Hubbard chains, we suddenly switched on the tunnel coupling along these chains and measured the emerging dynamics in terms of quasi-local densities, currents and coherences. Both the initial state preparation and the detection was realized using a bichromatic optical superlattice\ws\cite{Sebby:2006,Foelling:2007a}. For a wide range of (repulsive) inter-particle interactions, we find a fast relaxation of the measured observables to steady state values which are consistent with a dynamical version of Jaynes' principle\ws\cite{Jaynes:1957}. The timescale of the relaxation cannot be attributed to a classical ensemble average. For short times, we compare the experimental results to time-dependent density-matrix renormalization group simulations ($t$-DMRG, for a review see Refs.~\cite{Schollwoeck:2005, Schollwoeck:2011} and references therein) of the Hamiltonian dynamics without free parameters, further developing the ideas of previous numerical studies\ws\cite{Cramer:2008b, Flesch:2008}.

\paragraph*{Concept of the experiments.} 
We consider a one-dimensional chain of lattice sites coupled by a tunnel coupling $J$ and filled with repulsively interacting bosonic particles. In the tight-binding approximation, the Hamiltonian takes the form of a one-dimensional  Bose-Hubbard model\ws\cite{Jaksch:1998,Bloch:2008}
\begin{eqnarray*}
  \hat H &=& \sum_j \biggl[-J\left(\hat a^\dagger_j \hat a^{\phantom{\dagger}}_{j+1} +\mbox{h.c.}\right)
     +\frac{U}{2}   \hat n_j(\hat n_j -1) + \frac{K}{2}  \hat n_j j^2\biggr]\,,
\end{eqnarray*}
where $\hat a_{j}$ annihilates a particle on site $j$, $\hat n_{j} = \hat a_{j}^\dagger \hat a^{\phantom{\dagger}}_{j}$ reflects the number of atoms on site $j$ and $U$ is the on-site interaction energy. The parameter $K = m \omega^2 d^2$ ($m$ is the particle mass, $d$ the lattice spacing) describes an external harmonic trap with trapping frequency $\omega \simeq 2\pi \times 61\,{\rm Hz}$, present in the experiments. 

The experimental sequence can be described in three parts (see Fig.~1a): ({\it i}) At $t=0$, the system is initialized in a density wave represented as a state vector  $|\psi(t=0)\rangle = | \cdots ,1,0,1,0,1 ,\cdots\rangle$, such that only lattice sites with an even site index are occupied and no tunnel-coupling is present along the chain.  ({\it ii}) After the quench to a distinct set of positive parameters $J$, $U$ and $K$, the system follows the non-equilibrium dynamics of the above Hamiltonian $ \hat H$. ({\it iii}) Finally, the tunnel-coupling is suppressed again and the properties of the evolved state vector $|\psi(t)\rangle$ are read out.

% --- FIGURE 1 ---
\ifthenelse {\boolean{NP}}{}{
\begin{figure}[tb]\begin{center}
\includegraphics[width=0.4\textwidth]{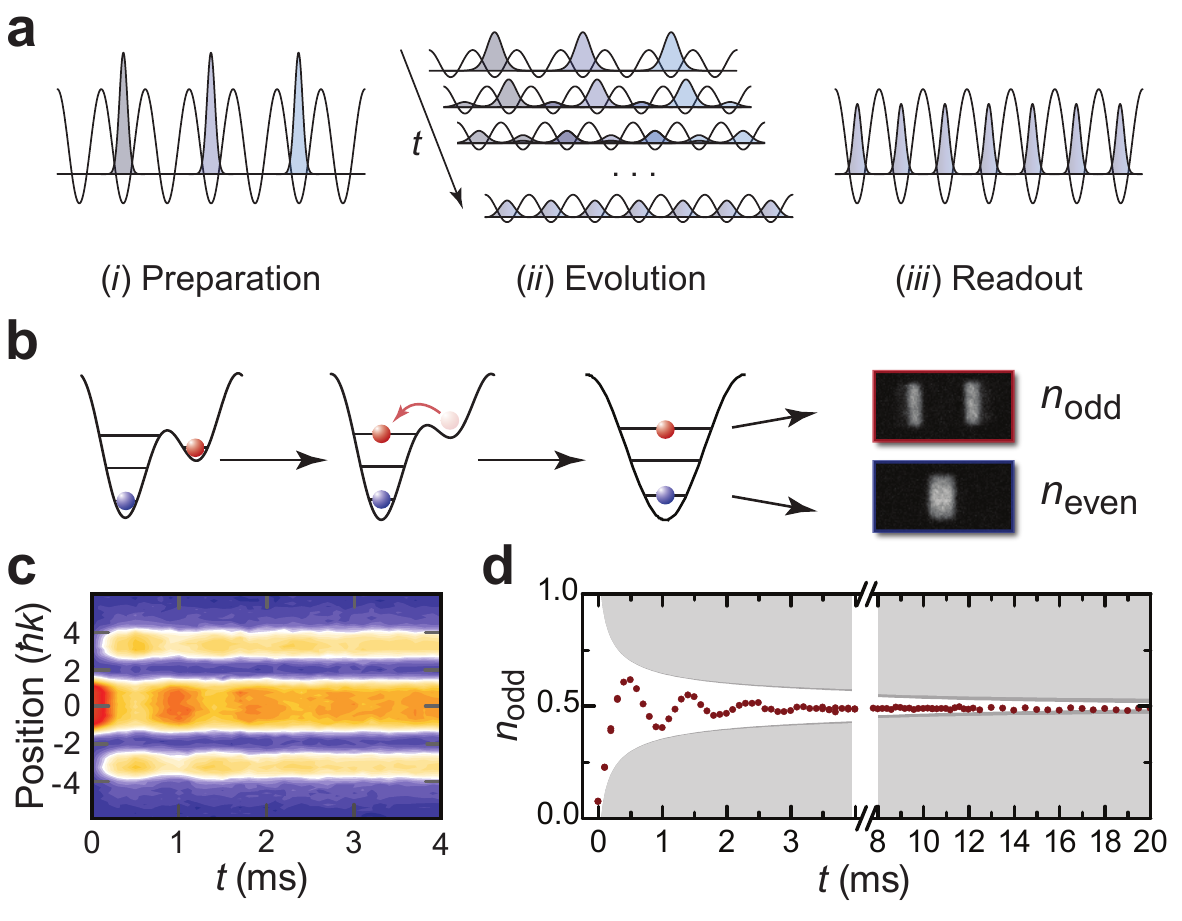}
\caption{\captionOne}
\end{center}\end{figure}
}

We started our experiments by loading a BEC of about $45\times 10^3$ $^{87}$Rb atoms in the $|F=1,m_F=-1\rangle$ Zeeman level into a 3D optical lattice formed by retroreflected laser beams of wavelength $\lambda_{xl} = 1530\,{\rm nm}$ along one direction (``long lattice") and $\lambda_{y,z} = 844\,{\rm nm}$ along the other two. In this loading we were crossing the transition to a Mott-insulator which resulted in an occupation of not more than one particle per site. Finally we added to the long lattice another optical lattice with wavelength $\lambda_{xs} = 765\,{\rm nm} = \lambda_{xl}/2$ (``short lattice") with the relative phase between the two adjusted to load every second site of the short lattice\ws\cite{Peil:2003,Foelling:2007a}. Completely removing the long lattice gave an array of practically isolated 1D density waves $|\psi_N\rangle = |\cdots ,1,0,1,0,1, \cdots \rangle$ -- thus realizing step ({\it i}) -- with a distribution of particle numbers $N$ and thus lengths $L = 2N-1$ given by the external confinement. For our parameters, we expect chains with a maximal particle number of $N_{\rm max} \simeq 43$ and a mean value of $\bar N \simeq 31$ (see Supplementary Material for details on the loading procedure).

To initialize the many-body relaxation dynamics of step ({\it ii}), we quenched the short-lattice depth to a small value within $200\,{\rm \mu s}$, allowing the atoms to tunnel along the $x$-direction. After a time $t$, we rapidly ramped up the short lattice to its original depth, thus suppressing all tunneling. Finally, we read out the properties of the evolved state in terms of densities, currents and coherences in step ({\it iii}). Note that in the experiments we always measured the full ensemble average $X(t) = \EE_{\{\!N\!\}} \langle\psi_N(t)| \hat X | \psi_N(t)\rangle$ of an observable $\hat X$ over the array of chains (denoted by the averaging operator $\EE_{\{\!N\!\}}$), rather than the expectation value for a single chain with $N$ particles.

\paragraph*{Relaxation of quasi-local densities.} 
We first discuss measurements of the density on sites with either even or odd index. After the time evolution, we transferred the population on odd sites to a higher Bloch band using the superlattice and detected these excitations employing a band-mapping technique (see Fig.~1b)\ws\cite{Sebby:2006,Foelling:2007a}. Fig.~1c shows the integrated band-mapping profiles as a function of relaxation time for $h/(4J) \simeq 0.9\,{\rm ms}$, $U/J = 5.16(7)$ and $K/J \simeq 9 \times 10^{-3}$. We plot the resulting traces $n_{\rm odd}(t)$ in Fig.~1d. We generally observe oscillations in $n_{\rm odd}$ with a period $T \simeq h/(4J)$ which rapidly dampen out within 3-4 periods to a steady value of  $\simeq 0.5$. The same qualitative behavior is found in a wide range of interactions (see Fig.~2).

% --- FIGURE 2 ---
\ifthenelse {\boolean{NP}}{}{
\begin{figure}[tb]
\begin{center}
\includegraphics[width=0.5\textwidth]{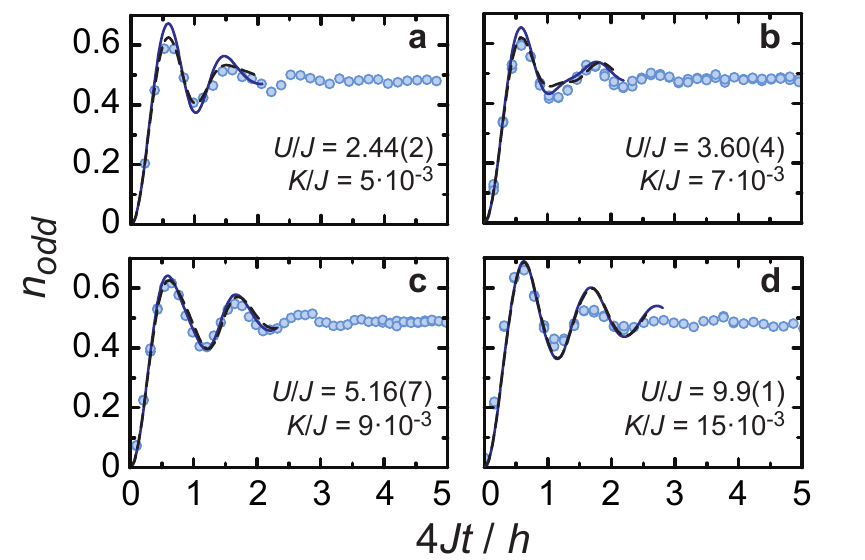}
\end{center}
\vspace{-0.5cm}
\caption{\captionTwo}
\end{figure}
}

We performed $t$-DMRG calculations, keeping up to $5000$ states in the matrix-product state simulations (solid lines in Fig.~2). The Bose-Hubbard parameters used in these simulations were obtained from the respective set of experimental control parameters. Furthermore, we took into account the geometry of the experimental setup by performing the corresponding ensemble average $\EE_{\{\!N\!\}}$ over chains with different particle numbers $N$ (see Supplementary Material). For the times accessible in the simulations, these averages differ only slightly from the traces obtained for a single chain with the maximal particle number $N_{\rm max} = 43$ of the ensemble (see Supplementary Material). For interaction strengths $U/J \lesssim 6$ (Fig.~2a-c), we find a good agreement of the experimental data and the simulations. In this regime, only small systematic deviations can be observed, which are strongest for the smallest value of $U/J$ which corresponds to the smallest lattice depth. 
They can be attributed to the breakdown of the tight-binding approximation for shallow lattices which gives rise to a significant amount of longer-ranged hopping. When including a next-nearest neighbor hopping term $-J_{\rm NNN}\sum_j (\hat a^\dagger_j \hat a_{j+2} + {\rm h.c.})$ in the $t$-DMRG simulations we obtain quantitative agreement with the experimental data (dashed line in Fig.~2).
For larger values of $U/J$ and correspondingly deeper lattices, the tight-binding approximation is valid. For $U/J \gtrsim 10$ (Fig.~2d), larger deviations are found. Here, the dynamics become more and more affected by residual inter-chain tunneling and non-adiabatic heating as the absolute timescale of the intra-chain tunneling $\propto 1/J$ becomes larger.

The results of the density measurements can be related to the expectations for an infinite chain with $K=0$. There, the time-evolution can be calculated analytically in the case of either non-interacting bosons ($U/J=0$) or infinite interactions ($U/J \to \infty$)\ws\cite{Cramer:2008b,Flesch:2008}. These limiting cases can be well-understood through the mechanism of local relaxation by ballistically propagating excitations. The on-site densities follow $0$-th order Bessel functions describing oscillations which are asymptotically dampened by a power law with exponent $-1/2$. The damping we observe in the interacting system, however, is much faster. This behavior has also been found in $t$-DMRG simulations of homogeneous Hubbard chains with finite interactions\ws\cite{Cramer:2008b,Flesch:2008}. The exact origin of this enhanced relaxation in the presence of strong correlations constitutes one of the major open problems posed by the results presented here.

\paragraph*{Measurements of quasi-local currents.}
Employing the bichromatic superlattice, we were also able to detect the magnitude and direction of quasi-local density currents. Instead of raising the short lattice at the end of step ({\it ii}), we ramped up the long lattice to suppress the tunnel-couplings through every second potential barrier in the chain (see Fig.~3a). At the same time, we set the short lattice to a fixed value to obtain always the same value of $(U/J)_{\rm DW} \simeq 0.2$ in the emerging double wells. By tuning the relative phase between the long and short lattice we were able to selectively couple sites with index $(2j,2j\!+\!1)$ (``even-odd", $j$ integer) or $(2j\!-\!1,2j)$ (``odd-even"). We recorded the time-evolution in the now isolated double-wells using the same final read-out scheme as for the densities (see Fig.~3b). We find sinusoidal tunnel oscillations which dephase only slowly and decrease in amplitude with increasing relaxation time $t$. The phase $\phi$ and amplitude $A$ of these oscillations were extracted from a fit of a sine-wave to the data and are plotted in Fig.~3c as a function of the relaxation time for $U/J = 5.16(7)$. While the phase contains the information about the direction of the mass flow, the amplitude is a combination of the local population imbalance and the strength of the local current.

% --- FIGURE 3 ---
\ifthenelse {\boolean{NP}}{}{
\begin{figure}[tb]
\begin{center}
\includegraphics[width=0.4\textwidth]{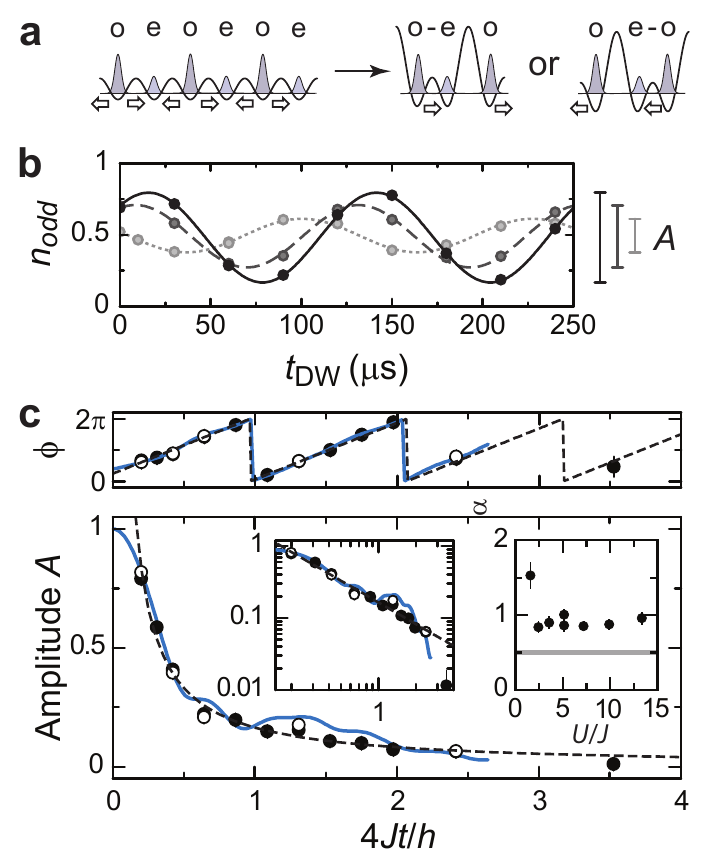}
\end{center}
\caption{\captionThree}
\end{figure}
}

We find $\phi$ to evolve linearly in time, giving strong evidence that the excitations in the system expand approximately ballistically as suggested in Refs.~\cite{Cramer:2008b,Flesch:2008}. Furthermore, its value does not change when coupling even-odd or odd-even sites, indicating the absence of center-of-mass motion in the system. The amplitude $A$ on the other hand decays to zero on the same timescale as the oscillations in the local densities dampen out -- in fact the quantities $(1\pm A)/2$ provide envelopes to the traces $n_{\rm odd}$ and $n_{\rm even}$ (see Supplementary Material). On short timescales, $0 < 4Jt/h < 3$, we find the decay of the amplitude to follow an approximate power-law $\propto t^{-\alpha}$ with $\alpha = 0.86(7)$. This behavior might well change at longer times, where no significant amplitude was measurable. We extract the power-law coefficients $\alpha$ for a wide range of $U/J$ (right inset to Fig.~3c). In all cases, the absolute values of the coefficients are larger than the one expected for free particles, where $\alpha = 0.5$, again indicating the faster relaxation in the presence of interactions. 

It is key to the experiment that the observed fast damping cannot be attributed to a mere classical ensemble averaging due to the inhomogeneous distribution of tunnel-couplings in the various chains ($\text{var}(J)/J \simeq 0.4\%$) or the external trap. Furthermore, we ensure that the transverse tunnel-coupling between adjacent chains $J_\perp$ is always one to two orders of magnitude smaller than $J$. Furthermore, the dynamics of a single site -- or of the densities of odd sites -- cannot be described in terms of simple rate equations, and not even in terms of Markovian quantum master equations reflecting damped motion (see Supplementary Material). Similarly, no dynamical mean-field description can capture the dynamics for large $U$~\cite{Hastings:2008}. Hence, any realistic description has to necessarily include the many-body and non-Markovian features of the dynamics, contributing to the challenge for a numerical simulation for intermediate times.

\paragraph*{Time-evolution of the quasi-momentum distribution.}
A different view on the relaxation can be obtained from the quasi-momentum distribution of the ensemble. When instantaneously switching off all trapping potentials after a relaxation time $t$ and letting the cloud expand freely for a time $t_{\rm ToF}$, the density distribution takes the form $n_{\rm ToF}(\vec r) \propto |\widetilde{w}_0(m\vec r/\hbar t)|^2 \mathcal S(m \vec r/\hbar t)$. Here, $\tilde{w}_0(\vec k)$ is the Fourier transform of the on-site Wannier orbital and the interference term for the ensemble of decoupled Hubbard chains in the far-field limit is $ \mathcal{S}(\vec k) = \EE_{\{\!N\!\}} \sum_{j,j'} e^{i k_x (j-j') d} \langle \hat a^\dagger_j \hat a^{\phantom{x}}_{j'} \rangle$ with $d = \lambda_{xs}/2$ being the lattice spacing along the chain direction. In Fig.~4a, we plot the measured density profiles integrated over the $y$- and $z$-direction as a function of the relaxation time (left panel) together with the corresponding patterns reconstructed from $t$-DMRG simulations for the full distribution of chains (right panel) for $U/J \simeq 5$. Both the experimental data and the numerical calculation show a rapid build-up of short-range coherence, not present in the initial state. 

% --- FIGURE 4 ---
\ifthenelse {\boolean{NP}}{}{
\begin{figure}[tb]
\begin{center}
\includegraphics[width=0.4\textwidth]{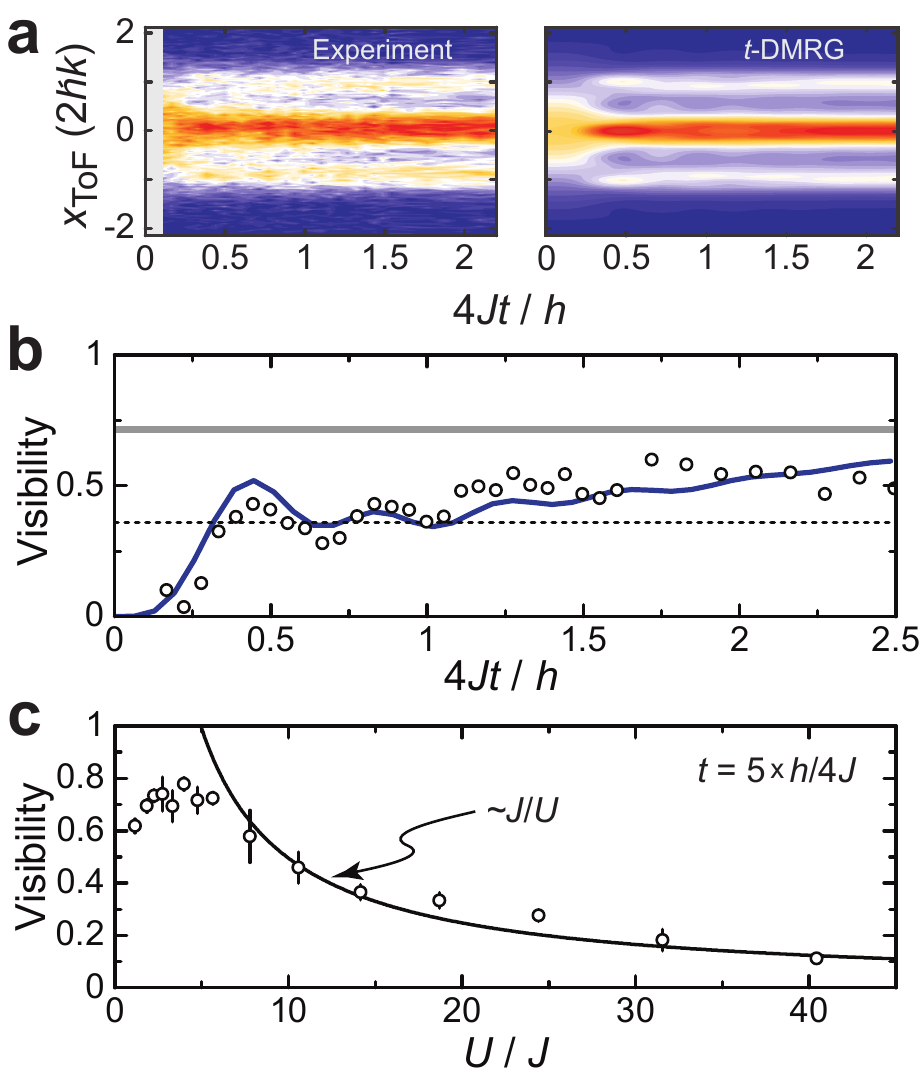}
\end{center}
\vspace{-0.5cm}
\caption{\captionFour}
\end{figure}
}

At short relaxation times $4Jt/h \lesssim 2$, the simulation data shows a strong cosinusoidal component with a period of $2 \hbar k = 2 ht_{\rm ToF}/(m\lambda_{xs})$ and weaker contributions from higher harmonics. While the former correspond to next-neighbor coherences in the system, the latter are a signature of correspondingly longer range coherences which rapidly decay in the relaxation process\ws\cite{Flesch:2008}. Due to the noise on the experimental data, the higher frequency components are weak, but can still be identified. For longer relaxation times $4Jt/h \gtrsim 2$, only the next-neighbor coherences remain as also found from $t$-DMRG simulations of homogeneous Hubbard chains with finite interactions\ws\cite{Flesch:2008}. We extract the visibility of the lowest frequency component as described in Ref.~\cite{Foelling:2007a} both from the experimental data and the $t$-DMRG calculations (see Fig.~4b). In the absence of local currents in the system, the visibility of the lowest Fourier component is given by $4 \EE_{\{\!N\!\}} \Re \langle \hat a^\dagger_j \hat a^{\phantom{x}}_{j+1}\rangle$. We find good agreement between experiment and numerics. The visibility builds up towards a first maximum at $4Jt/h \simeq 0.5$ corresponding to the first maximum in $n_{\rm odd}(t)$ (Fig.~2c), followed by dampened oscillations. From $4Jt/h\simeq 2$ on, we find the visibility to increase slowly with $t$ towards an equilibrium value which is significantly higher than the one found for a homogeneous system (dashed line in Fig.~4b). In the presence of the trap, the system can expand after the quantum quench, converting kinetic energy proportional to $ -\langle \hat a^\dagger_j \hat a^{\phantom{x}}_{j+1}\rangle$ into potential energy associated with
$ \langle \hat n_j j^2 \rangle$. As a consequence, the absolute value of $\langle \hat a^\dagger_j \hat a^{\phantom{x}}_{j+1}\rangle$ increases.

We observe the same qualitative behavior for the whole range of interaction strength $U/J$ accessed in the experiments, but with a strength of the next-neighbor correlations which depends on $U/J$. In Fig.~4c, we plot the visibility of the interference patterns found from experiments at different values of $U/J$ with at a relaxation time of $4Jt/h = 5$, where densities and currents are fully relaxed. We find the visibility to have a maximum around $U/J=4$ while it decreases towards the analytic limits of $U \to 0$ and $U \to \infty$. In both of these limits, the problem can be mapped onto free particles where no coherences will survive for long times. In the latter case of hardcore bosons, the coherences are found to be suppressed with $J/U$ in the experiments. The same behavior can be found from perturbation theory in $J/U$ for a thermal state in the lattice\ws\cite{Flesch:2008}. Recent theoretical results show that for non-degenerate spectra, the time-averaged state is always identical to the maximum entropy state given by the full set of constants of motion\ws\cite{Gogolin:2010}. On the other hand, if a state relaxes locally, it has to be to the projection of this time-averaged state on the respective subsystem. However, it is up to now unclear how to physically interpret and identify all constants of motion defining the maximum entropy state. Interestingly, the findings presented here are compatible with the expectations for a Gibbs state defined only by the total energy and the total number of particles. 
Finally note that the build-up of short-range phase coherence to a finite value complements the observed decay of the density pattern and the currents. It cannot be explained by any classical dephasing mechanism, but is a result of genuine many-body dynamics in the system.

For the time-evolution of the densities, currents and coherences, $t$-DMRG simulations and experimental results show a remarkable congruence. This emphasizes the clean implementation of controlled quantum dynamics in the one-dimensional interacting Hubbard-model and the high fidelity of the initial-state preparation. All parameters are calculated \textit{ab inito} from the experimental control parameters. Therefore, the experiments can be seen as a self-sustained dynamical quantum simulation where the simulation effort is the same for each value of $t$ and each set of parameters. As long as the time-evolution is not perturbed by experimental imperfection, this dynamical quantum simulator outperforms any continuous-time numerical simulation for which the calculational effort increases with $t$. Simulation methods on classical computers such as matrix-product state based time-dependent DMRG used here suffer from an extensive increase in entanglement entropy which limits the relaxation times accessible in the calculations\ws\cite{Calabrese:2006b,Osborne:2006}.

\paragraph*{Conclusions and Outlook.}
In conclusion, we have demonstrated measurements on the relaxation of a charge-density wave of in a strongly correlated one-dimensional Bose gas with varying interactions. Using a bichromatic optical superlattice, we were able to prepare a patterned density state with high fidelity and to induce non-equilibrium dynamics by rapidly switching on the tunnel-coupling along the chain. We could follow the dynamical evolution of the initial state in terms of even- and odd-site densities, local currents and short-range correlations visible in the quasi-momentum distribution. We have compared our measurements to parameter-free $t$-DMRG simulations, finding excellent agreement and identifying contributions from next-nearest neighbor hopping. All three observables can be seen as local probes of the system and show a rapid relaxation to steady state values as it is predicted for Hubbard-type models by a central limit theorem\ws\cite{Cramer:2008a, Cramer:2010}. These steady state values are compatible with the system globally being in a maximum entropy state. This idea can be seen as a dynamical version of Jaynes' principle which could recently be substantiated theoretically for local observables or reduced states\ws\cite{Linden:2009, Cramer:2008b, Barthel:2008} and for two-periodic observables\ws\cite{Flesch:2008, Hastings:2008} as considered here. After a finite time, the closed quantum system cannot be distinguished locally from having reached a global Gibbs state under the constraint of a set of macroscopic constants of motion set by the initial state\ws\cite{Jaynes:1957, Rigol:2007, Barthel:2008,Cramer:2008a}.

A direct measurement of global observables to, e.g., identify constants of motion of the dynamics is inhibited by the ensemble of chains with various particle numbers in the experimental realization. This limitation might be overcome by either preparing a single chain with fixed length or by selecting a single chain from the ensemble for detection. This will open the way to answer what global state is after all reached in the evolution, including possible pre-thermalization\ws\cite{Moeckel:2008} and experimental studies of the eigenstate thermalization hypothesis\ws\cite{Deutsch:1991,Srednicki:1994,Rigol:2008}. It is the hope that the present work triggers further experimental studies which address theoretically unsolved key questions of non-equilibrium dynamics.

% ---- METHODS SECTION ----
\ifthenelse {\boolean{NP}}
{
\begin{methods}
% One single supplementary materials appendix for the arxiv version
\end{methods}

\paragraph*{References}
}{\thankyou}

%%%%% BIBLIOGRAPHY
%\bibliography{Relaxation}

%merlin.mbs 2010-03-15 4.21a (PWD, AO, DPC)

%%%%% BIBLIOGRAPHY-END

\ifthenelse {\boolean{NP}}
{\noindent \textbf{Acknowledgments}

\noindent \thankyou

\noindent \textbf{Author Contributions}

\noindent \contributions

\noindent \textbf{Additional information}

\noindent \declare

}{}

\ifthenelse {\boolean{NP}}{
\ifthenelse {\boolean{FIG}}{
\clearpage
\begin{figure}[p]\begin{center}
\includegraphics[width=0.7\textwidth]{Fig1}
\end{center}
\caption{\captionOne}
\end{figure}

\begin{figure}[p]\begin{center}
\includegraphics[width=0.8\textwidth]{Fig2}
\end{center}
\caption{\captionTwo}
\end{figure}

\begin{figure}[p]\begin{center}
\includegraphics[width=0.6\textwidth]{Fig3}
\end{center}
\caption{\captionThree}
\end{figure}

\begin{figure}[p]\begin{center}
\includegraphics[width=0.6\textwidth]{Fig4}
\end{center}
\caption{\captionFour}
\end{figure}
}{
\clearpage

\noindent\textbf{Figure 1 $\vert$ }\captionOne
\bigskip

\noindent\textbf{Figure 2 $\vert$ }\captionTwo
\bigskip

\noindent\textbf{Figure 3 $\vert$ }\captionThree
\bigskip

\noindent\textbf{Figure 4 $\vert$ }\captionFour
}
}{

\clearpage
\section*{Supplementary Material}

\section{Loading procedure and sudden quench}
We started the loading procedure by ramping up the long lattice to a depth of $30\,{E_r^{xl}}$ which results in an array of isolated two-dimensional quantum gases. The lattice depth is given in units of the respective recoil energy $E_r^{i} = h^2/(2m\lambda_i^2)$. Subsequently, we ramped up the two transverse optical lattices with wavelengths of $\lambda_{y,z} = 844\,{\rm nm}$ to $30\,E_r^{y,z}$, crossing the Mott insulator transition in each of the two-dimensional gases with a maximal filling of one atom per site (see Supplementary Material).  We then added the short lattice with a depth of $30\,{E_r^{xs}}$ to the long lattice, forming a bichromatic period-two superlattice\ws\cite{Foelling:2007a}. We set the relative phase between the short and the long lattice such that every second short-lattice site coincided with a minimum of the long-lattice potential. Finally removing the long lattice completely yielded an ensemble of one-dimensional density waves $|\cdots ,1,0,1,0,1, \cdots \rangle$~\cite{Foelling:2007a,Sebby:2007a} with various particle numbers $N$.

We verified the loading of single atoms by bringing each atom into a superposition of the states $|F=1,m_F=-1\rangle$ and $F=2,m_F=1\rangle$, where atom pairs can undergo hyperfine-relaxing collisions which would expel them from the lattice. We measured the total number of particles with ($N_{\rm tot}$) and without this filtering step ($N_{n=1}$), finding that up to a total number of $60 \times 10^3$ atoms, no pairs were removed from the lattice (see Fig.~\ref{fig:SuppFig1}a). For the experiments, we chose a particle number of $45\times 10^3$ in order to obtain maximally singly occupied sites.

\begin{figure}[tb]
\begin{center}
\includegraphics[width=0.45\textwidth]{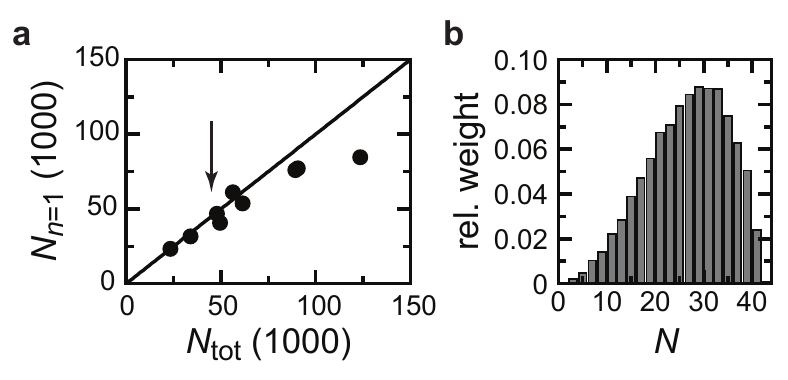}
\end{center}
\vspace{-0.5cm}
\caption{\textbf{a} Measured atom number on sites with $n=1$ versus the total atom number. The solid line represents the condition $N_{\rm tot} = N_{n=1}$. The arrow indicates the total atom number used in our experiments. \textbf{b} Distribution of particle numbers per chain caused by the harmonic confinement calculated for our loading sequence.}\label{fig:SuppFig1}
\end{figure}

\section{DMRG simulation respecting the geometry of the problem}
After the loading procedure, the occupied sites form an ellipsoid around the center of the trap. Thus, due to the suppressed transversal hopping, the experimental setup is described by a two-dimensional approximately circular array of one-dimensional chains with different particle numbers $N$ with $N_{\textrm{max}}=43$. For a realistic theoretical description, DMRG calculations were performed for several particle numbers ranging from $5$ to $43$ and weighted according to the distribution shown in Fig.~\ref{fig:SuppFig1}b. In Fig.~\ref{fig:SuppFig15}, we plot the averaged densities as obtained from DMRG and shown in Fig.~2 together with the results for a single chain with the maximal number of particles. Both curves differ only slightly on the timescales accessible by the simulations.

In the experiment, the smooth external trap leads to a slow expansion of the cloud along the $x$-direction, rather than a sharp reflection of excitations traveling at a velocity $\propto J$ in the system. Thus, the length of the one-dimensional lattices in the DMRG calculations ($\sim$ 121) was chosen such that the particles do not reach the system boundary during the simulated times. This absence of a sharp reflection also explains, together with the averaging over different chains with different particle numbers for which recurrences would happen at different times, why no recurrences of the density wave are visible in our experiments.

\begin{figure}[tb]
\begin{center}
\includegraphics[width=0.45\textwidth]{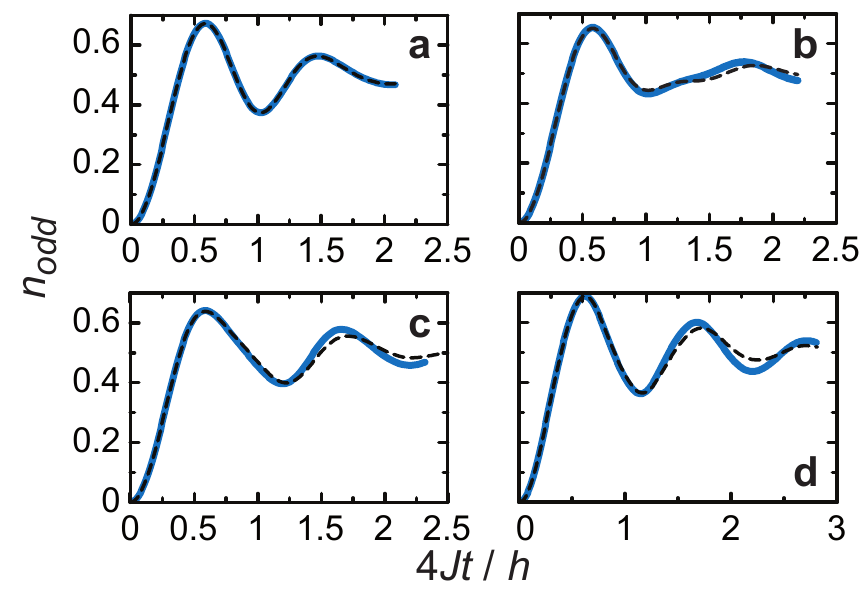}
\end{center}
\vspace{-0.5cm}
\caption{DMRG results for the relaxation of the quasi-local densities $n_{\rm odd}$. The solid lines represent the ensemble averaged $t$-DMRG data plotted in Fig.~2a-d. The dashed lines are the simulation results for a single chain in the ensemble with $N=43$ particles. The parameters used in the simulations are the same as in Fig.~2.}\label{fig:SuppFig15}
\end{figure}

\section{Current measurements}
We measure the quasi-local currents by suppressing every second tunnel coupling in $200\,{\rm \mu s}$ with the help of the superlattice, such as to obtain chains of symmetric double-wells (see Fig.~3a). For all current measurements, we chose the superlattice depths to be $40\,E_r^{xl}$ and $4\,E_r^{xs}$, respectively. In this configuration, we have a tunnel coupling within the double-wells of $J_{\rm DW} \simeq h\times 8\,{\rm kHz}$ and an interaction strength of $(U/J)_{\rm DW} \simeq 0.2$, such that the particles in the double-well will practically tunnel independently. In Fig.~3b, we plot the relative population on the left side of the double wells $n_L$ as a function of the holdtime $t_{\rm DW}$ and for three different relaxation times as they were evaluated for Fig.~3c. Here, we have measured the population on the left and right sides of the double wells by the same band mapping technique which was used for the density measurements. The phase $\phi$ and the amplitude of the oscillations shown in Fig.~3c are extracted from a simple sinusoidal fit to the data.

If we assume the particles in the double-well formed by the sites with index $j$ and $j+1$ to evolve under the simple single-particle Hamiltonian $\hat H_{{\rm DW},j} = -J_{\rm DW} (\hat a^\dagger_j \hat a_{j+1} + {\rm h.c.})$, we find the oscillation amplitude to be
\begin{equation}
  A_j(t) = \sqrt{\big(\langle \hat n_{j}(t) \rangle - \langle \hat n_{j+1}(t) \rangle\big)^2 
  + 4\, \Im\big(\langle \hat a_j(t)^\dagger \hat a_{j+1}^{\phantom{\dagger}}(t)\rangle\big)^2}\,,
\end{equation}
where $t$ is the relaxation time. From the simultaneous measurement of the local densities, it is thus possible to reconstruct the bare mass current $\Im\langle \hat a_j(t)^\dagger \hat a_{j+1}{\phantom{\dagger}}(t)\rangle$ through the barrier between the two sites as a function of the relaxation time $t$. Furthermore, it is evident from Eq.~(1), that whenever the mass current vanishes the amplitude is measuring the population imbalance between the sites $j$ and $j+1$. At these points, the quantities $(1 \pm A)/2$ reproduce the even- and odd-site densities, respectively. Therefore they provide two envelopes to $n_{\rm odd,even}(t)$.

For the phase $\phi_j(t)$, we find from the same calculation as above
\begin{equation}
  \phi_j(t) = \arctan\left(
    -\frac{2\Im\big(\langle \hat a_j(t)^\dagger \hat a_{j+1}^{\phantom{\dagger}}(t)\rangle\big)}
    {\langle \hat n_{j}(t) \rangle - \langle \hat n_{j+1}(t) }\right)\,.
\end{equation}
We use these two relations to calculate the ensemble-averaged amplitude 
$\EE_{\{N\}} A(t)  =  \vert \EE_{\{N\}} \sum_j A_i(t) \exp(i \phi_j(t)) \vert$ 
and phase 
$\EE_{\{N\}} \phi(t)  =  {\rm arg}[ \EE_{\{N\}} \sum_j A_i(t) \exp(i \phi_j(t)) ]$ 
from the DMRG simulations (solid lines in Fig.~3b).

\section{Deviation from Markovian and mean-field dynamics}
In this section, we show that the time evolution cannot be described by a Markovian quantum master equation, signifying the complex relaxation dynamics beyond a situation that can be described in terms of rates. That is to say, each constituent does not see the rest
of a chain as a mere bath it is weakly coupled to, but intricate memory effects do play a role.
To see this, it is sufficient to consider the 
simple case of local relaxation dynamics of a single site and the case of $U=0$. We restrict ourselves
to the infinite translationally invariant case and an initial state of a density wave as described above---any 
other setting is only more complex and in general also non-Markovian.
To follow a Markovian time evolution of the reduced state of some odd site 
$j$, denoted as $\rho$, means that  its time evolution follows a master equation in Lindblad form
\begin{eqnarray}
  \frac{d}{d t} \rho(t) 
 &=& i[\rho(t), \hat h] + \sum_{\alpha,\beta} G_{\alpha,\beta}\Bigl(
   \hat F_\alpha\rho(t)  \hat F_\beta^\dagger
  - \frac{1}{2}\{ \hat F_\beta^\dagger  \hat F_\alpha,\rho(t)\}_+
  \Bigr),\nonumber
\end{eqnarray}
where $\hat h$ is a Hermitian operator that can be different from the free Hamiltonian $\hat H$, 
and $G$ and $\hat F_\alpha$ are some arbitrary matrices, reflecting the influence of the rest of the chain. 
This is the most general form of a master equation when the dynamical map, mapping the initial state 
$|1\rangle\langle1|=\rho(0)\mapsto \rho(t)$, reflects Markovian dynamics. 
Since this is a bosonic free model, one can keep track of the evolution
by specifying the first and second moments alone. The second moment matrix of
$\rho(t)$  is given by, with $\hat b=\hat b_j$, $\hat X=(\hat b+\hat b^\dagger)/\sqrt{2}$ and $P=i(\hat b^\dagger -\hat b)/\sqrt{2}$,
\begin{equation*}
	\gamma(t)=\left[
	\begin{array}{cc}
		2\langle \hat X^2\rangle (t)&\langle \hat X \hat P \rangle (t) + \langle \hat P \hat X\rangle(t)\\
		\langle \hat X \hat P \rangle (t) + \langle \hat  P \hat X\rangle(t) & 2 \langle \hat P^2\rangle(t)
	\end{array}
	\right],
\end{equation*}
and the evolution is given by the  Gaussian channel $\gamma(t) = A(t)\gamma(0)A^T(t)+B(t)$. A straightforward
calculation gives
\begin{eqnarray*}
	B(t)=\biggl( \sum_{k=1}^{L/2}  (2|V_{j,2k}(t) |^2   +1) \biggr) \id_2,\,\,
	A(t) =  |V_{j,j}(t) |^2  \id_2,
\end{eqnarray*}	
with $V(t)=e^{-itH}$, as a matrix exponential. 

This evolution then has to be compared with a Markovian time evolution, one that satisfies
\begin{equation*}
	A(s+s') = A(s) A(s'),\,\, B(s+s') = A(s) B(s') A(s)^T + B(s) 
\end{equation*}	
as would necessarily
be true for a dynamical semi-group generated by a Gaussian Markovian master equation.
A reasonable figure of merit of Markovianity of a Gaussian channel, is, for a given time $t\geq 0$, how
different the dynamical map is from a Markovian one. There exists more than one reasonable way to 
quantify such a difference, yet all lead to qualitatively identical results. A physically well-defined way of 
quantifying this is a norm difference from the Jamiolkowski covariance matrix of the closest Markovian 
Gaussian channel to the given channel at hand \cite{Wolf:2008}. A more pragmatic---but in the certification of
non-Markovianity equally effective---means is,  for a time $t\geq 0$, to simply compute
\begin{equation}
	\inf \|A(s+s') - A(s)A(s')\|_2,
\end{equation}
with the minimization being performed with respect to $s,s'\geq 0$ such that $s+s'=t$. This quantity 
can easily be seen to take positive values for intermediate times, giving rise to 
a bound that 
can be directly related to observable differences on states.
Hence, no Markovian dynamics---amounting to rate equations---can model the dynamics encountered
here, and memory and genuine quantum many-body effects have to be considered.

The dynamics found here is also not consistent with a mean-field picture for values of $U$ significantly 
different from
zero. In such a picture, 
one looks for a time-dependent 
self-consistent solution for
\begin{equation*}
	x(t) = \frac{1}{L}\sum_{j} (-1)^{j-1} \langle \hat n_j \rangle (t),
\end{equation*}
for $L$ sites
 in a mean-field Hamiltonian
\begin{equation*}
	\hat{H}(t) =-\sum_{j}\left[
	\hat{b}^\dagger_{j+1}\hat{b}_j+ 
	\hat{b}^\dagger_{j}\hat{b}_{j+1} +
	\frac{U}{2} \bigl(- \frac{1}{2} +(-1)^{j-1} x(t)\bigr) \hat{n}_j
	\right]\,,
\end{equation*}  
in a variant of the findings of Ref.\ \cite{Hastings:2008}.
Using Runge-Kutta numerical integration one finds that although for short times and small $U$, the
evolution of densities $\langle \hat n_j\rangle(t)$ is quite compatible with a mean-field picture, one encounters 
significant deviations
for larger $U$ and larger times. This again shows the non-triviality of the dynamics, in that a mean field picture cannot
capture the dynamics at hand, and sophisticated $t$-DMRG simulations are necessary.

}

\end{document}